\documentstyle[aaspp4]{article}

\def\stacksymbols #1#2#3#4{\def\theguybelow{#2}
        \def\verticalposition{\lower#3pt}
        \def\spacingwithinsymbol{\baselineskip0pt\lineskip#4pt}
        \mathrel{\mathpalette\intermediary#1}}
\def\intermediary #1#2{\verticalposition\vbox{\spacingwithinsymbol
        \everycr={}\tabskip0pt
        \halign{$\mathsurround0pt#1\hfil##\hfil$\crcr#2\crcr
                \theguybelow\crcr}}}

\catcode`\@=11
\def\gsim{\ifmmode{\mathrel{\mathpalette\@versim>}}
    \else{$\mathrel{\mathpalette\@versim>}$}\fi}
\def\lsim{\ifmmode{\mathrel{\mathpalette\@versim<}}
    \else{$\mathrel{\mathpalette\@versim<}$}\fi}
\def\@versim#1#2{\lower 2.9truept \vbox{\baselineskip 0pt \lineskip 
    0.5truept \ialign{$\m@th#1\hfil##\hfil$\crcr#2\crcr\sim\crcr}}}
\catcode`\@=12

\def\msol{M_\odot}

\def\Mbh{M_{\rm BH}}
\def\Mbhz{M_{\rm BH,0}}
\def\Mbhu{M_{\rm BH,1}}

\def\Re{R_{\rm e}}
\def\Rez{R_{{\rm e},0}}

\def\sigc{\sigma_{\rm c}}
\def\sigcz{\sigma_{{\rm c},0}}
\def\sigcu{\sigma_{{\rm c},1}}
\def\sigcd{\sigma_{{\rm c},2}}
\def\sigct{\sigma_{{\rm c},3}}

\def\Lz{L_0}
\def\Lu{L_1}
\def\Ld{L_2}
\def\Lt{L_3}

\def\no{n_1}
\def\nd{n_2}

\slugcomment{accepted version - March 19, 2001}

\lefthead{Ciotti \& van Albada}
\righthead{The FP and $\Mbh -\sigc$ relation}

\begin{document}

\title{The $\Mbh -\sigc$ relation as a constraint on the formation 
       of elliptical galaxies}
\author{Luca Ciotti\altaffilmark{1,3,4} and 
        Tjeerd S. van Albada\altaffilmark{2}}
\affil{$^1$Department of Theoretical Physics, 1 Keble Road, Oxford, 
OX1 3NP, UK\\
ciotti@bo.astro.it}
\affil{$^2$Kapteyn Institute, Groningen University, 
The Netherlands\\
albada@astro.rug.nl}

\altaffiltext{3}{On leave from Osservatorio Astronomico di Bologna, 
via Ranzani 1, I-40127 Bologna, Italy}
\altaffiltext{4}{Also Scuola Normale Superiore, Piazza dei Cavalieri 7,
I-56126 Pisa, Italy}


\begin{abstract}

Most elliptical galaxies contain central black holes, whose masses
scale proportionally to the observed central velocity dispersions of
the host galaxies according to the so--called $\Mbh -\sigc$
relation. Here we discuss some consequences that can be derived by
combining the $\Mbh -\sigc$ relation with the scaling relation
describing the Fundamental Plane of elliptical galaxies.  In
particular, the possibility of substantial dissipationless merging in
the formation and evolution of elliptical galaxies is discussed.
Enforcing the merger end--products to satisfy the two scaling
relations mentioned above, a major role of dissipation in galaxy
formation is strongly suggested by our analysis. Moreover, we show
that existing observational data may shed some light on the complex
process of black hole merging.

\end{abstract}



%

\section{Introduction}


In recent years the presence of massive black holes (BHs) in the
centers of elliptical galaxies has been firmly established (for a
review see, e.g., de Zeeuw 2000). The black hole masses ($\Mbh$) lie
in the range $10^6 - 10^9\msol$ and correlate surprisingly well with
the stellar velocity dispersions in the central regions of the host
galaxies:
\begin{equation}
\Mbh\propto\sigc^{\alpha}.
\end{equation}
The precise value of the slope $\alpha$ of this relation is still
somewhat uncertain and appears to depend on the choice of galaxies and
on the fitting procedure. For example, Gebhardt et al.  (2000,
hereafter G00) find $\alpha= 3.75\pm 0.3$, whereas Ferrarese \&
Merritt (2000, hereafter FM00) find $\alpha=4.8\pm 0.5$; a discussion
of the possible origin of this difference can be found in Merritt \&
Ferrarese (2001).

The existence of a relation between BH mass and stellar velocity
dispersion may well provide valuable information on the process of
galaxy formation, and some of its implications have already been
discussed by Haehnelt \& Kauffmann (2000, hereafter HK00).  In fact,
two substantially different scenarios for the formation of elliptical
galaxies have been proposed. In the monolithic collapse picture
ellipticals are formed at early times by dissipative processes (see,
e.g., Eggen, Lynden--Bell, \& Sandage 1962; Larson 1974), while in
cold dark matter cosmologies elliptical galaxies result from a process
of hierarchical merging. In the latter scenario the assembly of
massive ellipticals requires several mergers, the last major merger
taking place in relatively recent times, i.e. at $z<1$ (see, e.g.,
White \& Rees 1978, Kauffmann 1996). This latter view is supported by
evidence of substantial galaxy merging in the cluster MS 1054-03 (van
Dokkum et al. 1999). The data for this cluster show that the galaxies
involved were typically E/S0 or early--type S in the pre--merger
stage.

Here we explore the consequences of combining equation (1) with other
scaling relations followed by elliptical galaxies, such as the
Faber--Jackson relation (1976, hereafter FJ) and the Fundamental Plane
(FP) relation between $\Re$, $L$, and $\sigc$ (Dressler et al. 1987;
Djorgovski \& Davis 1987), where $\Re$ is the effective radius, $L$
the total luminosity (for example in the Johnson B-band), and $\sigc$
the central velocity dispersion. Taking the point of view that
early-type galaxies mainly form through merging and that the $\Mbh -
\sigc$ and FP relations are preserved (or produced) by merging, we
argue that 1) either the appropriate rule for adding the mass of
merging BHs is substantially different from what is commonly assumed,
or 2) the merging process involves a significant dissipative phase.
The requirement for significant dissipation has been recognized
already on other grounds (see, e.g., Kauffmann \& Haehnelt 2000, HK00,
Burkert \& Silk 2000).

\section{Scaling relations and the dissipationless merging hypothesis}

For simplicity, let us assume that the primordial building blocks of
elliptical galaxies were stellar systems characterized by $(\Lz, \Rez,
\Mbhz, \sigcz)$, and that during dissipationless merging the galaxy
luminosities are added linearly. Then, by assembling $\no$ primordial
galaxies, we obtain $\Lu=\no\Lz$, and equation (1) predicts that the
central velocity dispersion $\sigcu$ of galaxy $\Lu$ correlates with
$\Mbhu$, the mass of the BH resulting from the merger.

Unfortunately, how BH masses combine when they merge is not well
known: energy (mass) can be radiated during merging as gravitational
waves and the total amount of energy radiated in a merger of two black
holes is highly uncertain, because detailed calculations of these
mergers are not yet available (see, e.g., Flanagan \& Hughes 1998,
hereafter FH98, Centrella 2000).  However, using the entropy--area
relation for BHs, it can be shown that in the maximally efficient
radiative merging (hereafter ``ME''), the square of the final mass
equals the sum of the squares of the initial masses. This is because
$S_{\rm BH}\propto A_{\rm BH}\propto\Mbh^2$ (where $S_{\rm BH}$ and
$A_{\rm BH}$ are the BH entropy and surface area, respectively), and
in the ME case the final area is equal to the sum of the inital BH
areas (Hawking 1976, see also Peacock 1999).  To proceed we assume
\begin{equation}
\Mbhu^{\mu}=\no\Mbhz^{\mu},
\end{equation}
where $0<\mu <2$ is a free parameter. Thus, the ME and classical cases
(i.e., when no energy is emitted as gravitational waves) are
recovered, respectively for $\mu =2$ and $\mu=1$.  With $\mu\leq 1$ we
can mimic in a simple way the case in which a significant amount of
orbital kinetic energy (compared to $\no\Mbhz c^2$) is available just
before the BH merging process and is converted into mass, resulting in
a final BH with a mass greater than the linear sum of the BH masses
involved in the merger\footnote{In fact, given $a_i\geq 0$
($i=1,...,n$) and two numbers $q\geq p\geq 0$, then $(\sum
a_i^p)^{1/p}\geq (\sum a_i^q)^{1/q}$.}. Note, however, that work done
so far on equal mass BH mergers indicates (in our notation) $\mu\gsim
1.4$ (FH98, and references therein), and that $\mu <1$ seems
physically unlikely.

From equations (1) and (2) it follows that $\sigcu/\sigcz
=\no^{1/\mu\alpha}$, therefore $\Lu/\Lz=(\sigcu/\sigcz)^{\mu\alpha}$.
Let us now assume that another galaxy is assembled by the merging of
$\nd$ primordial objects, and that the galaxies ``1'' and ``2'' also
merge together in a second generation of mergers, and so on.  It is
trivial to show that if $\Lt=\Lu+\Ld$ than
$\Lt/\Lu=(\sigct/\sigcu)^{\mu\alpha}$ and
$\Lt/\Ld=(\sigct/\sigcd)^{\mu\alpha}$.  The same argument can be
repeated for all successive merging events.  Thus, {\it if the total
luminosity is conserved during a merger, if $\mu$ is independent of
the BH masses, and if the $\Mbh -\sigc$ relation is preserved, then
the end products satisfy the relations}
\begin{equation}
\Mbh\propto L^{1/\mu}, 
\end{equation}
\begin{equation}
L\propto\sigc^{\mu\alpha}.
\end{equation}
Note that the treatment described is unaffected by the presence of
dark matter halos.  The first scaling relation above is just the
Magorrian et al. (1998) $\Mbh -L$ relation if one adopts $\mu\simeq
1/1.2\simeq 0.8$. The second relation is the FJ relation provided
$\mu\alpha\simeq 4$. By using $\mu=0.8$ one obtains $\alpha=5$, a
value compatible with that derived observationally by FM00.  If
instead of $\alpha=4.8$ one uses $\alpha=3.75$ (G00), then the FJ
relation can be reproduced by assuming $\mu\simeq 1.1$, but then there
is a conflict with the observed $\Mbh -L$ relation. The $\Mbh -L$
relation is characterized by a substantial scatter however, of at
least one order of magnitude, and so its constraining power is not as
strong as that of the FJ relation. Thus, according to the above
analysis, {\it dissipationless merging as the main way to assemble
massive ellipticals could be a viable solution (marginally) consistent
with the FJ, $\Mbh -L$, and $\Mbh -\sigc$ scaling relations, provided
one is willing to exclude maximally efficient radiative (or even
classical) merging of BH's and $\alpha$ is high}.

Of course, in the idealized scenario here explored we cannot reproduce
the scatter of the relations above, because the seed galaxies are
assumed to be identical in {\it all} their properties. Possible
sources of scatter in the two scaling relations given in equations (3)
and (4) could be differences in the properties of the objects
belonging to the ``zero-th generation'', and variations of the $\mu$
value depending on the details of the merging process.

Fortunately, we can carry our analysis further by using a
significantly tighter correlation, i.e., the Fundamental Plane
relation.  In the following we require the end--products of the
mergers to lie on the FP, for which we choose the relation obtained by
J{\o}rgensen, Franx \& Kj{\ae}rgaard (1996) in the B-band for a large
sample of galaxies:
\begin{equation}
\Re\;\sigc^{1.82}\;L^{-1.26}=constant,
\end{equation}
where $\sigc$ is the central velocity dispersion corrected to a
circular aperture with radius $\Re/8$.  The exponents in the equation
above depend (slightly) on the wave band used for the observations,
but as will be clear from the following discussion, these differences
do not affect our conclusions.  Although the FP and the FJ relations
both describe a property of the distribution of elliptical galaxies in
$\Re$, $L$, $\sigc$ space, they are essentially independent. The FP
relation defines a plane in $\Re$, $L$, $\sigc$ space containing the
large majority of elliptical galaxies, whereas the FJ relation
describes a projection of the distribution of galaxies within that
plane.

From equations (4) and (5), the following scaling relation should be
satisfied by galaxies formed through dissipationless merging:
\begin{equation}
\Re\propto \sigc^{1.26\mu\alpha -1.82}.
\end{equation}
Note that the quantity $\sigc$ appearing in equations (1) and (5) is
defined in the same way only in the case of assuming the FM00 slope
(in their analysis FM00 use the observed velocity dispersion inside a
circular aperture of radius $\Re/8$, while G00 adopt the average
velocity dispersion inside the effective radius, $\sigma_{\rm e}$).
However, in the following we discuss also the case of the G00 slope,
mainly for two reasons: the first is to investigate the degree of
robustness of our results with respect to possible changes in the
value of $\alpha$, and, second, because Merritt \& Ferrarese (2001)
showed that the mean value of the ratio $\sigma_{\rm e}/\sigc$ (for
their sample of 27 galaxies and with the exception of the Milky Way),
is 1.01.

To compare the prediction in equation (6) with observations we
considered three galaxy samples, obtained by combining the data (in
the Johnson B-band) from Bender, Burstein \& Faber (1992, 87 objects
in their Table 1, excluding dwarf spheroidals), and from J{\o}rgensen,
Franx \& Kj{\ae}rgaard (1995ab, 83 galaxies from the clusters Abell
194, Abell 3574, S 753, DC2345-28).  We used galaxies with $\sigc\geq
100$ km s$^{-1}$ only, in order to restrict our analysis to galaxies
(except two cases) for which the central BHs have secure mass
estimates (see Table 1 in FM00, and Table 1 in Merritt \& Ferrarese
2001). In particular, sample I, containing 170 galaxies, is the
combination of the samples of Bender et al. and J{\o}rgensen et
al. described above.  In sample II (151 galaxies), 19 bulges from
Bender et al. in sample I have been excluded. Sample III (83
galaxies), consists of the galaxies from J{\o}rgensen et al. .

With the standard least-squares method (see, e.g., Bevington 1969), we
fitted the two parameters of the linear relation
\begin{equation}
\log\Re=a+b\log\sigc,
\end{equation}
where $\Re$ is given in kpc and $\sigc$ in km s$^{-1}$. The results
for the three samples are given in Table 1; they agree within the
errors.

In agreement with the preceding discussion based on the $\Mbh -L$ and
FJ relations, the main result obtained by using the FP is that the
assumption of ME (or classical) BH merging cannot be reconciled with
the observations.  The above conclusion may be illustrated by
estimating upper and lower limits for the predicted slope
$b=1.26\mu\alpha-1.82$ (see equation [6]) from the observed value of
$\alpha$ and the cases of ME and classical BH merging. This yields
respectively $b\simeq 10.3$ ($\mu = 2$ and $\alpha =4.8$) and $b\simeq
2.9$ ($\mu =1$ and $\alpha = 3.75$), to be compared with the observed
value $b\simeq 1$ (Table 1).  Another way to present this
inconsistency is to determine the expected $\mu$ value from the last
column in Table 1: adopting $\alpha=3.75$ we obtain $\mu\simeq 0.6$,
while for $\alpha=4.8$ $\mu\simeq 0.5$.

As well known, the dependence of the FP coefficients on the
photometric band is weak and, therefore, the argument above is not
affected by the choice of the wave band: for example, the {\it
minimum} $b$ value obtained (when $\mu=1$ and $\alpha=3.75$) using the
FP coefficients given in Table 5 of J{\o}rgensen et al. (1996) is 2.7.

\placefigure{fig1}

In Figure 1 we present the $(\log\Re,\log\sigc)$ plane for sample I.
The data show a large dispersion around the best fit (heavy solid
line). The lines representing the {\it expected} $\Re,\sigc$
correlations in case of ME and classical BH merging for the two
adopted values of $\alpha$ are well above the data points however: in
other words the expected effective radii of massive galaxies are far
too large.

\section{Discussion and Conclusions}

The inconsistency between the $\Mbh -\sigc$ relation and
dissipationless hierarchical merging with $\mu\geq 1$ one the hand,
and the FP on the other hand, can be explained in physical terms as
follows.  The number of mergers required to increase $\sigc$ by a
factor $r$ is $n =r^{\mu\alpha}$. Thus, an increase of a factor $r$ in
the central velocity dispersion corresponds to an increase of
$r^{\mu\alpha}$ in the luminosity. If the resulting galaxies are
required to lie on the FP relation, then $\Re$ must increase
considerably due to the increase in luminosity (see equation [5]). In
other words, {\it the basic reason of the large increase in the
effective radius is the large number of merging events required to
increase the central velocity dispersion}\footnote{Note that from the
virial theorem and energy conservation the {\it virial} velocity
dispersion in dissipationless galaxy merging cannot increase.  In this
scenario the increase of $\sigc$ with $\Mbh$ must be due to
non--homology effects. But observational evidences points to dynamical
and structural homology in elliptical galaxies (see, e.g., Gerhard et
al. 2001) and thus provides a strong argument against dissipationless
merging.}. HK00 suggest that ``frequent merging of galaxies in
hierarchical cosmogonies moves galaxies along the correlation in the
$\Mbh-\sigc$ plane, even when galaxies are gas--poor and their BHs
grow mainly by the merging of pre--existing BHs'': it would be of
great interest to test if in this case the ellipticals resulting from
the mergers actually lie on the observed FP, and at the appropriate
location.

Following the general discussion above, three possibilities can be
suggested that may solve this inconsistency.

1. In the case of purely dissipationless merging, the addition of BH
masses in a merger is neither the result of ME, nor classical merging:
in other words, a (fine tuned) significant fraction of the orbital
kinetic energy of the merging BHs (just before the merging event) is
converted into mass. In this case $\mu$ is decreased below unity,
leading to agreement with the results of Section 2. However, in the
commonly accepted scenario in which the BHs merge, dissipating their
orbital kinetic energy and dynamically heating the stellar background
of the newly formed galaxy (see, e.g., Faber et al. 1997), it is not
clear whether the amount of extra--kinetic energy required is
available.

2. ME or classical merging of black holes may be a viable possibility
in case gas is present in the merging galaxies. In fact, a gaseous
component could affect the merging in two ways.  First, the central BH
in the merger remnant may accrete gas. Equation (2) is then no longer
valid: a source term must be added on the r.h.s. .  This situation can
be mimicked with a decreased {\it effective} $\mu$ in equation (2),
possibly to the small values ($\mu\simeq 0.5$) that are apparently
required by the data, in combination with ME or classical merging of
the BH's in the progenitor galaxies. Second, the formation of stars
from a centrally condensed gas component in the merger remnant might
lead to a decrease of the - overall - effective radius.  Of course, in
both cases fine--tuning between the amount of gaseous
accretion/dissipation and the masses of the merging BHs and the parent
galaxies is needed because the $\Mbh-\sigc$ and FP relations must be
maintained. In addition to this difficulty, it has been pointed out
that BH mergers which radiate too much energy anisotropically may also
radiate a substantial amount of linear momentum (see, e.g., FH98), and
the consequent recoil of the final BH could correspond to a kick
velocity that is a moderate fraction of the speed of light, sufficient
to expell the resulting BH from the galaxy. If this were confirmed, ME
merging can perhaps not be reconciled with the merging scenario for
ellipticals, even in presence of a substantial gaseous dissipation.

3. Finally, the more likely possibility is that the bulk of
ellipticals is not formed by merging at late times, but as a
consequence of strongly dissipative processes at high redshift. In
this case the $\Mbh -\sigc$ correlation (as well as the
metallicity--velocity dispersion relation) are the product of the
complicated feedback processes associated with galaxy formation (see,
e.g., Ostriker 1980, Burkert \& Silk 2000).

\acknowledgments

L.C. is indebted to Giuseppe Bertin, James Binney, Alessandro
Braccesi, Alberto Cappi, Giacomo Giampieri, Jerry Ostriker, Silvia
Pellegrini, Alvio Renzini, Renzo Sancisi, and Scott Tremaine for
discussions, and to the warm hospitality of Merton College in Oxford
(UK).  T.v.A. is grateful to Bologna Astronomical Observatory for its
hospitality.  L.C. was supported by MURST, contract CoFin2000.  We
thank the anonymous referee for constructive remarks.
\clearpage
\begin{deluxetable}{crrrrrrrr}
\footnotesize
\tablecaption{\label{tbl-1}}
\tablewidth{0pt}
\tablehead{
           \colhead{Sample} & \colhead{$N$} & 
                              \colhead{$a\pm\sigma_a$} & 
                              \colhead{$b\pm\sigma_b$} &
                              \colhead{$\chi^2$} &
                              \colhead{$\mu\alpha$} &
          }
\startdata

I                   &         170           &   -1.86$\pm$ 0.53  &
                         1.14$\pm$ 0.23     &   0.15             &
                         2.35\\

II                  &         151           &   -1.34$\pm$ 0.51  &
                         0.94$\pm$ 0.22     &   0.13             &
                         2.19\\

III                 &          83           &   -1.49$\pm$ 0.70  & 
                         1.06$\pm$ 0.31     &   0.12             &
                         2.29\\

\enddata \tablecomments{The parameters $a$ and $b$ of equation (5),
with their errors $\sigma_a$ and $\sigma_b$, and the reduced $\chi^2$
of the fit for the three data samples. $N$ is the number of galaxies
in each sample. In the last column we give the product $\mu\alpha$
required in equation (6) to reproduce observations.}
\end{deluxetable}
\clearpage


%
%

\clearpage
\figcaption[]{The (logarithmic) $(\Re,\sigc)$ plane for our sample I
(170 galaxies). The heavy line is the best fit ($b=1.14$), the other
lines are equation (5) evaluated with the $(\mu,\alpha)$ pairs given
in the figure.
\label{fig1}} 
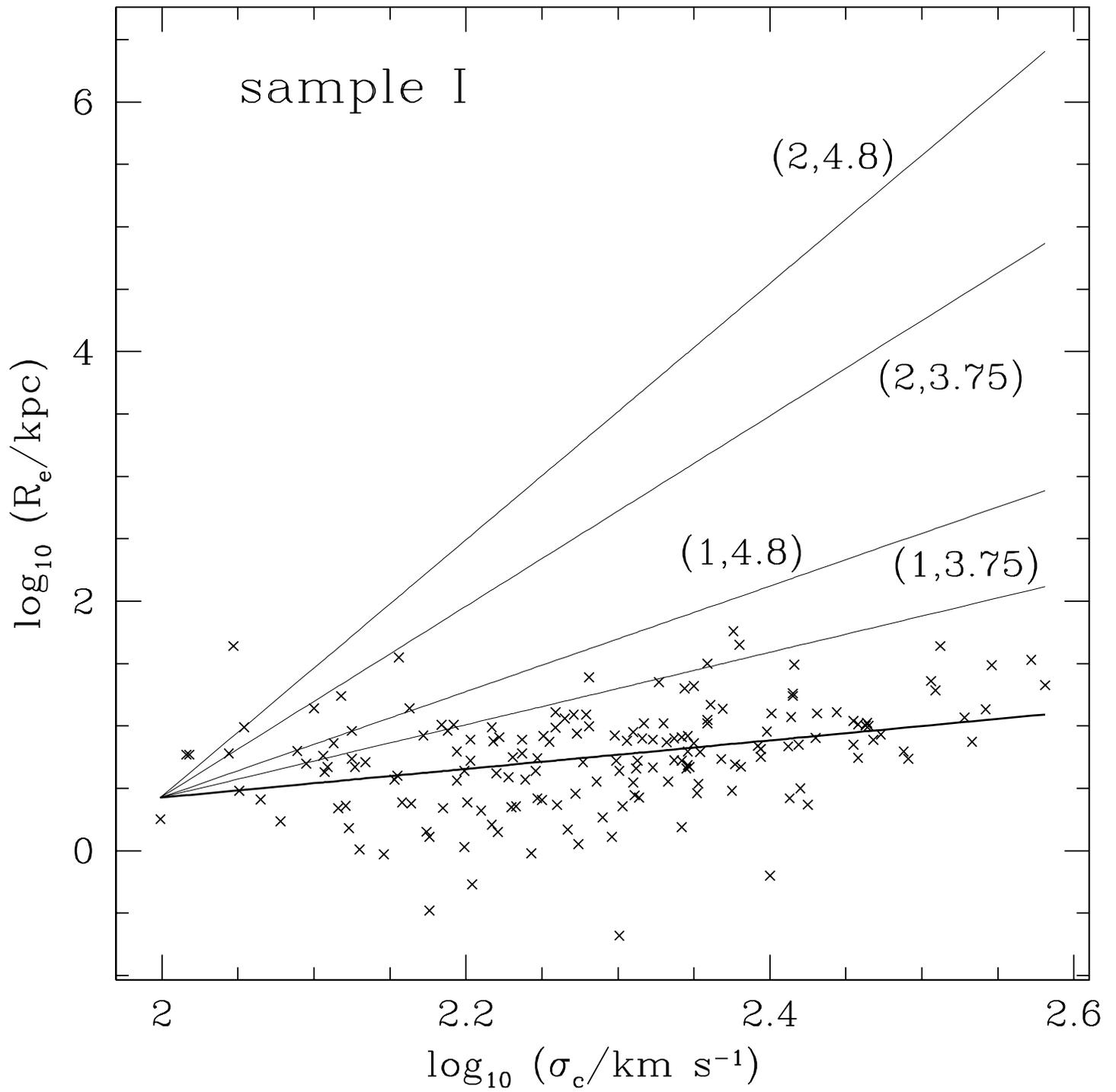
\end{document}